\documentclass[letterpaper, 10 pt, conference]{ieeeconf}%
\usepackage[utf8]{inputenc}
\usepackage[T1]{fontenc}
\usepackage{amsmath}
\usepackage{graphicx}
\usepackage{cite}
\usepackage{bm}
\usepackage{amsfonts}
\usepackage{threeparttable}
\usepackage{multirow}
\usepackage{algorithm}
\usepackage{algorithmic}
\usepackage{booktabs}
\usepackage{amssymb}
\usepackage{color}%
\providecommand{\U}[1]{\protect\rule{.1in}{.1in}}
\IEEEoverridecommandlockouts
\newtheorem{problem}{\textbf{Problem}}

\begin{document}

\title{{\LARGE \textbf{A General Framework for Decentralized Safe Optimal Control of
Connected and Automated Vehicles in Multi-Lane Intersections }}}
\author{Huile Xu, Wei Xiao, Christos G. Cassandras,~\IEEEmembership{Fellow,~IEEE,} Yi Zhang,~\IEEEmembership{Member,~IEEE,} and Li Li,~\IEEEmembership{Fellow,~IEEE}
\thanks{The work of H. Xu, W. Xiao, and C. G. Cassandras was supported in part by NSF under grants ECCS-1509084, DMS-1664644,
CNS-1645681, IIS-1723995, CPS-1446151, by AFOSR under grant FA9550-19-1-0158,
by ARPA-E's NEXTCAR program under grant DE-AR0000796 and by the MathWorks. The work of H. Xu, L. Li, and Y. Zhang was supported in part by the National Key Research and Development Program of China under Grant 2018YFB1600600.}
\thanks{H. Xu is with the Department of Automation, BNRist, Tsinghua
University, Beijing 100084, China and also with the Division of Systems
Engineering and Center for Information and Systems Engineering, Boston
University, Brookline, MA, 02446, USA. (e-mail:
hl-xu16@mails.tsinghua.edu.cn)} \thanks{W. Xiao and C. G. Cassandras are with
the Division of Systems Engineering and Center for Information and Systems
Engineering, Boston University, Brookline, MA, 02446, USA. (Email: \{xiaowei,
cgc\}@bu.edu) } \thanks{Y. Zhang is with Department of Automation, BNRist,
Tsinghua University, Beijing 100084, China and also with the Tsinghua-Berkeley
Shenzhen Institute (TBSI), Tower C2, Nanshan Intelligence Park 1001, Xueyuan
Blvd., Nanshan District, Shenzhen 518055, China. (Email: zhyi@tsinghua.edu.cn)
} \thanks{L. Li is with the Department of Automation, BNRist, Tsinghua
University, Beijing 100084, China. (e-mail: li-li@tsinghua.edu.cn)}}
\maketitle

\begin{abstract}
We address the problem of optimally controlling Connected and Automated Vehicles (CAVs) arriving from four multi-lane roads at an intersection where they conflict in terms of safely crossing (including turns) with no collision. The objective is to jointly minimize the travel time and energy consumption of each CAV while ensuring safety. This problem was solved in prior work for single-lane roads. A direct extension to multiple lanes on each road is limited by the computational complexity required to obtain an explicit optimal control solution. Instead, we propose a general framework that first converts a multi-lane intersection problem into a decentralized optimal control problem
for each CAV with less conservative safety constraints than prior work. We then employ a method combining optimal control and control barrier functions, which has been shown to efficiently track tractable unconstrained optimal CAV trajectories while also guaranteeing the satisfaction of all constraints. Simulation examples are included to show the effectiveness of the proposed framework under symmetric and asymmetric intersection geometries and different schemes for CAV sequencing.
\end{abstract}


\thispagestyle{empty} \pagestyle{empty}



\section{INTRODUCTION}

Intersections are the main bottlenecks for urban traffic. As reported in
\cite{rios2016survey}, congestion in these areas causes US commuters to spend
6.9 billion hours more on the road and to purchase an extra 3.1 billion
gallons of fuel, resulting in a substantial economic loss to society. The
coordination and control problems at intersections are challenging in terms of
safety, traffic efficiency, and energy consumption
\cite{rios2016survey,chen2015cooperative}.

The emergence of Connected and Automated Vehicles (CAVs) provides a promising
way for better planning and controlling trajectories to reduce congestion and
ultimately improve safety as well as efficiency. Enabled by vehicle-to-vehicle
(V2V) and vehicle-to-infrastructure (V2I) communication, CAVs can exchange
real-time operational data with vehicles in their vicinity and communicate
with the infrastructure \cite{li2014survey}. Based on these technologies,
researchers have made significant progress in the optimal control of CAVs at
intersections. One of the prevailing ideas is to formulate an optimization
problem whose decision variables are both crossing sequences and control
inputs (e.g., desired velocity and desired acceleration). Hult \emph{et al.}
\cite{hult2016coordination} formulated the traffic coordination problem at a
three-way intersection as an optimal control problem where it is required that
the trajectories of any two vehicles do not intersect in order to guarantee
safety. However, it is not explicitly explained how to realize such safety
constraints. Some recent studies show that we can realize safety constraints
by introducing binary variables to represent crossing sequences, which results
in Mixed-Integer Linear Programming (MILP) problems \cite{fayazi2018mixed}.
Though some techniques, such as the grouping scheme \cite{xu2019grouping},
have been proposed to accelerate the computation process, it is still
difficult to extend this method to \emph{multi-lane} intersections with a
large number of vehicles which may also change lanes along the way. In terms
of optimizing control inputs, Model Predictive Control (MPC) is effective for
problems with simple (usually linear or linearized) vehicle dynamics and
constraints. However, when the vehicle dynamics are highly nonlinear and
complex, the whole problem becomes a non-linear MPC whose computation time is
still prohibitive for practical applications \cite{qian2015decentralized}.

Another idea is to decompose the whole optimization problem into two separate
problems, i.e., first determining the crossing sequence and then solving for
CAV control inputs according to this crossing sequence \cite{xu2020bi}. The most
straightforward crossing sequence mechanism is the First-In-First-Out (FIFO)
rule. For example, Dresner and Stone \cite{dresner2008multiagent} proposed
an autonomous intersection management cooperative driving strategy which
divides the intersection into grids (resources) and assigns these grids to
CAVs in a FIFO manner. In
\cite{malikopoulos2018decentralized} and
\cite{zhang2019decentralized}, a decentralized
optimal control framework is designed for CAVs to jointly minimize energy and
time by deriving the desired CAV arrival times at an intersection based on the
FIFO crossing sequence. However, some recent studies have shown that the
performance of the FIFO mechanism can lead to poor performance in some cases
\cite{meng2018analysis}. In \cite{zhang2018decentralized},
a Dynamic Resequencing (DR) scheme is designed to adjust the crossing sequence
whenever a new CAV enters the intersection control zone and showed that this
scheme is computationally efficient and improves traffic efficiency. Xu
\emph{et al.} \cite{xu2019cooperative} proposed a Monte Carlo Tree Search
(MCTS)-based cooperative strategy to find a promising crossing sequence for
all CAVs and demonstrated that this strategy may determine a good enough
sequence even for complicated multi-lane intersections where the search space
is enormous. After determining the crossing sequence, these studies also use
the analytical solutions proposed in \cite{malikopoulos2018decentralized} and
\cite{zhang2019decentralized} to solve for the optimal control inputs.
Although \cite{xiao2019decentralized} has extended this solution to consider
speed-dependent rear-end safety constraints, the resulting computational cost
significantly increases. Moreover, it is difficult to generalize this optimal
control method for complex and nonlinear vehicle dynamics without incurring
computational costs which make its real-time applicability prohibitive.

To address the above limitations, we propose an approach which combines the
use of Control Barrier Functions (CBFs) with the conventional optimal control
method to bridge the gap between optimal control solutions (which represent a
lower bound for the optimal achievable cost) and controllers which can
provably guarantee on-line safe execution \cite{Wei2020}. The key idea is to design CAV
trajectories which optimally track analytically tractable solutions of the
basic intersection-crossing optimization problem while also provably
guaranteeing that all safety constraints are satisfied. Through CBFs, we can
map continuously differentiable state constraints into new control-based
constraints. Due to the forward invariance of the associated set
\cite{ames2014control,ames2019control, Xiao2019}, a control input that
satisfies these new constraints is also guaranteed to satisfy the original
state constraints. This property makes the CBF method effective even when the
vehicle dynamics and constraints become complicated and include noise.

Along these lines, the main contribution of this paper is a novel
decentralized optimal control framework combining the optimal control and CBF
methods for a \emph{multi-lane} intersection. Specifically, we first formulate
the multi-lane intersection problem as an optimal control problem whose
objective is to jointly minimize the traffic delay and energy consumption
while guaranteeing that all CAVs safely cross a four-way intersection that
includes left and right turns. Unlike prior work, we replace roadway segments
referred to as \textquotedblleft merging zones\textquotedblright\ or
\textquotedblleft conflict zones\textquotedblright\ by \emph{Merging Points
(MPs)} which are much less conservative while still guaranteeing collision
avoidance. Allowing lane-changing behavior, we design a strategy to determine
the desirable locations of lane-changing MPs for all CAVs. Then, we propose a
search algorithm to quickly determine rear-end safety constraints and lateral
safety constraints that every CAV has to meet. Once these constraints are
specifed for any CAV, we design an \emph{Optimal Control and Barrier Function
(OCBF)} controller for solving the problem efficiently, as verified through
multiple simulation experiments. Our framework can accommodate a variety of
resequencing methods for finding a near-optimal crossing sequence, including
the aforementioned DR and MCTS schemes, leading to improved performance
compared to the FIFO rule, especially when the intersection is geometrically asymmetrical.

The paper is organized as follows. \emph{Section II} formulates the multi-lane
intersection problem as an optimal control problem with safety constraints
applied to a sequence of MPs. \emph{Section III} introduces our search
algorithm for determining all safety constraints pertaining to a given CAV,
and \emph{Section IV} presents our joint optimal and control barrier function
controller. \emph{Section V} validates the effectiveness of the proposed
method through simulation experiments. Finally, \emph{Section VI} gives
concluding remarks.

\section{PROBLEM FORMULATION}

\begin{figure*}[ptbh]
\centering
\includegraphics[width=17cm]{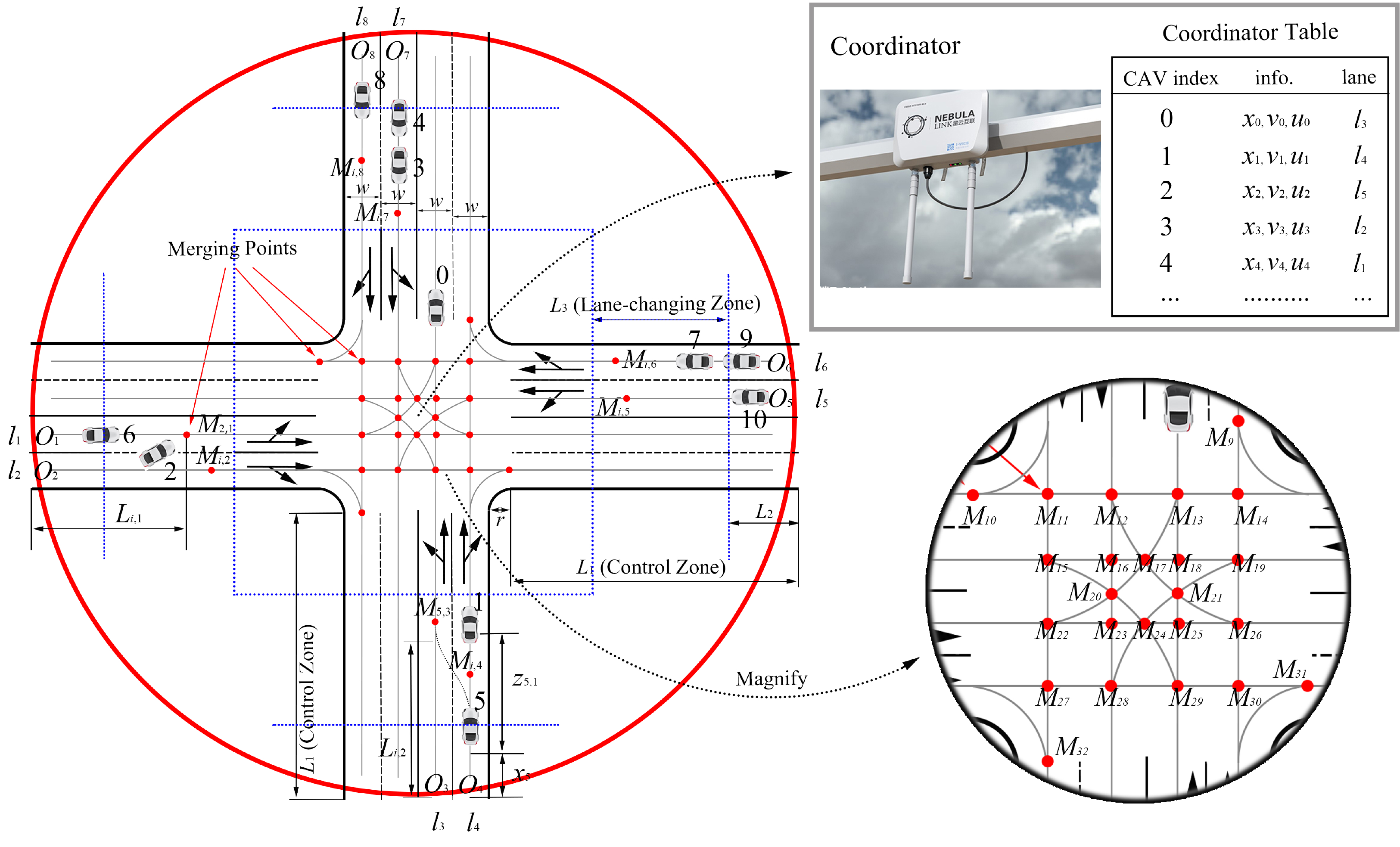}\caption{The multi-lane
intersection problem. Collisions may happen at the MPs (red dots shown in
above figure).}%
\label{fig:inter}%
\end{figure*}

Figure \ref{fig:inter} shows a typical intersection with multiple lanes. The area within the
outer red circle is called a \emph{Control Zone (CZ)}, and the length of each
CZ segment is $L_{1}$ which is initially assumed to be the same for all entry
points to the intersection; extensions to asymmetrical intersections are
straightforward and discussed in Section V.D. The significance of the CZ is
that it allows all CAVs to share information and be automatically controlled
while in it. Red dots show all MPs where potential collisions may occur. We
assume that the motion trajectory of each CAV in the intersection is
determined upon its entrance to the CZ (see grey lines in Fig. 1). Based on
these trajectories, all MPs in the intersection are fixed and can be easily
determined. However, unlike prior similar studies, we also allow possible
lane-changing behaviors in the CZ, which adds generality to our method. In
order to avoid potential collisions with new coming vehicles and to conform
with the driving rules that vehicles are prohibited from changing lanes when
they are very close to the intersection, these lane changes are only allowed
in a \textquotedblleft lane-changing zone\textquotedblright, i.e., an area
between the two blue lines shown in Fig. 1. We use dotted lines to
differentiate these zones from the rest of the area where solid lines are
shown. The distance from the entry of the CZ to the lane-changing zone is
$L_{2}$ and the length of the lane-changing zone is $L_{3}$. Since we
initially consider a symmetrical intersection, $L_{2}$ and $L_{3}$ are the
same for all lanes. However, it is easy to extend our method to asymmetrical
intersections and set different parameters for each lane, as shown in Section
V.D. Due to lane-changing, apart from the fixed MPs in the intersection, some
\textquotedblleft floating\textquotedblright\ MPs may also appear in
lane-changing zones which are not fixed in space.

We label the lanes from $l_{1}$ to $l_{8}$ in a counterclockwise direction
with corresponding origins $O_{1}$ to $O_{8}$. The rightmost lanes in each
direction only allow turning right or going straight, while the leftmost lanes
only allow turning left or going straight. However, all CAVs have three
possible actions: going straight, turning left, and turning right. Thus, some
CAVs must change their lanes so as to execute an action, e.g., left-turning
CAV $2$ in $l_{2}$ in Fig. 1. Due to such lane-changing behavior, a new MP
$M_{2,1}$ is generated since a conflict of CAV $2$ with a CAV in $l_{1}$ may
arise. Similarly, possible MPs may also appear in other lanes when vehicles
perform lane-changing maneuvers, as the red dots ($M_{i,2},M_{i,3},\cdots$,
and $M_{i,8}$) indicate in Fig. 1. Moreover, it is worth noting that if a CAV
needs to change lanes, then it has to travel an additional distance; we assume
that this extra distance is a constant $l>0$. In what follows, we consider an
intersection which has two lanes in each direction, one of the most common
intersection configurations worldwide, and observe that this model is easy to
generalize to intersections with more than two lanes.

The intersection has a coordinator (typically a Road Side Unit (RSU)) whose
function is to maintain the crossing sequence and all individual CAV
information. The most common crossing sequence is based on the FIFO queue of all CAVs, regardless of lanes, using their
arrival time at the CZ. The FIFO queue is fair and simple to implement,
however, its performance can occasionally be poor. Thus, various cooperative
driving strategies have been proposed to generate a more promising crossing
sequence, as in \cite{zhang2018decentralized, xu2019cooperative, Wei2020ACC}.
Our approach for controlling CAVs does not depend on the specific crossing
sequence selected. Therefore, we first use the FIFO queue so as to enable
accurate comparisons with related work also using this scheme and then
generalize it to include other resequencing methods which adjust the crossing
sequence whenever there is a new arriving vehicle, e,g., the DR method in
\cite{zhang2018decentralized}. This allows CAVs to overtake other CAVs in the
CZ from different roads, which better captures actual intersection traffic behaviors.



We begin by reviewing the same basic model as in previous work
\cite{zhang2019decentralized}, which will allow us to make accurate
comparisons. When a CAV enters the CZ, the coordinator will assign it a unique
index. Let $S(t)$ be the set of FIFO-ordered CAV indices and $N(t)$ be the
cardinality of $S(t)$. Based on $S(t)$, the coordinator stores and maintains
an information table, as shown in Fig. 1. For example, the current lane of CAV
$2$ changes from $l_{2}$ to $l_{1}$ after it completes a lane-changing
maneuver. In addition, after CAV $0$ passes the intersection, its index will
be dropped from the table and the indices of all other CAVs decrease by one.
This table enables each CAV to quickly identify CAVs that have potential
collisions with it and to optimize its trajectory to maximize some specific
objectives. The search algorithm for identifying conflicting vehicles will be
introduced in detail in the next section.

The vehicle dynamics for CAV $i$ take the form
\begin{equation}
\left[
\begin{array}
[c]{c}%
\dot{x}_{i}(t)\\
\dot{v}_{i}(t)
\end{array}
\right]  =\left[
\begin{array}
[c]{c}%
v_{i}(t)+w_{i,1}(t)\\
u_{i}(t)+w_{i,2}(t)
\end{array}
\right]  , \label{VehicleDynamics}%
\end{equation}
where $x_{i}(t)$ is the distance to its origin $O_{j}$ along the lane that CAV
$i$ is located in when it enters the CZ, $v_{i}(t)$ denotes the velocity, and
$u_{i}(t)$ denotes the control input (acceleration). Moreover, to compensate
for possible measurement noise and modeling inaccuracy, we use $w_{i,1}%
(t),w_{i,2}(t)$ to denote two random processes defined in an appropriate
probability space.

\subsection{Optimization Problem}

Based on the notation established above, we can now view trajectory planning of vehicles as
an optimization problem where we consider two objectives for each CAV subject
to three constraints, including the rear-end safety constraint with the
preceding vehicle in the same lane, the lateral safety constraints with
vehicles in the other lanes, and the vehicle physical constraints, as detailed next.

\textbf{Objective 1} (Minimize travel time): Let $t_{i}^{0}$ and $t_{i}^{f}$
denote the time that CAV $i$ arrives at the origin and the time that it
leaves the intersection, respectively. To improve traffic efficiency, we wish
to minimize the travel time $t_{i}^{f}-t_{i}^{0}$ for CAV $i$.

\textbf{Objective 2} (Minimize energy consumption): Apart from traffic
efficiency, another objective is energy efficiency. Ignoring any noise terms
in (\ref{VehicleDynamics}) for the time being, since energy consumption is a
monotonic function of the acceleration control input, an energy consumption
function we use is defined as
\begin{equation}
J_{i}=\int_{t_{i}^{0}}^{t_{i}^{f}}\mathcal{C}(u_{i}(t))dt, \label{eqn:obj}%
\end{equation}
\noindent where $\mathcal{C}(\cdot)$ is a strictly increasing function of its argument.

\textbf{Constraint 1} (Rear-end safety constraint): Let $i_{p}$ denote the
index of the CAV which physically immediately precedes $i$ in the CZ (if one
is present). To avoid rear-end collisions, we require that the spacing
$z_{i,i_{p}}(t)\equiv x_{i_{p}}(t)-x_{i}(t)$ be constrained by:
\begin{equation}
z_{i,i_{p}}(t)\geq\varphi v_{i}(t)+\delta,\text{ \ }\forall t\in\lbrack
t_{i}^{0},t_{i}^{m}], \label{Safety}%
\end{equation}
where $\delta$ is the minimum safety distance, and $\varphi$ denotes the
reaction time (as a rule, $\varphi=1.8s$ is suggested, e.g.,
\cite{vogel2003comparison}). If we define $z_{i,i_{p}}$ to be the distance
from the center of CAV $i$ to the center of CAV $i_{p}$, then $\delta$ is a
constant determined by the length of these two CAVs (thus, $\delta$ is
generally dependent on CAVs $i$ and $i_{p}$ but taken to be a constant over
all CAVs in the sequel, only for simplicity). Note that $i_{p}$ may change
when a lane change event or an overtaking event (discussed in Section III.B) occurs.

\textbf{Constraint 2} (Lateral safety constraint): Let $t_{i}^{k}$ denote the
time that CAV $i$ arrives at the MP $M_{k},k\in\{1,2,\dots,32\}$. CAV $i$ may
collide with other vehicles that travel through the same MP with it. For all
MPs, including the floating MPs $M_{i,l}$ due to lane-changing, there must be
enough safe space when CAV $i$ is passing through, that is,
\begin{equation}
z_{i,j}(t_{i}^{k})\geq\varphi v_{i}(t_{i}^{k})+\delta_{i}, \label{SafeMerging}%
\end{equation}
where CAV $j\neq i$ is a CAV that may collide with $i$ (note that $j$ may not
exist and that there may also be multiple CAVs indexed by $j$ for which this
constraint applies at different $t_{i}^{k}$). The determination of $j$ is
challenging, and will be addressed in the following section. Compared with
related work that requires no more than one CAV within a conflict (or merging)
zone at any time, we use (\ref{SafeMerging}) to replace this conservative
constraint. Instead of such a fixed zone, the space around the MPs to define collision avoidance
now depends on the CAV's speed (and possibly size if we allow $\delta_{i}$ to
be CAV-dependent), hence it is much more flexible.

\textbf{Constraint 3} (Vehicle physical limitations): Due to the physical
limitations on motors and actuators, there are physical constraints on the
velocity and control inputs for each CAV $i$:
\begin{equation}
\begin{aligned} v_{min} &\leq v_i(t)\leq v_{max}, \forall t\in[t_i^0,t_i^{m}],\\ u_{i,min}&\leq u_i(t)\leq u_{i,max}, \forall t\in[t_i^0,t_i^{m}], \end{aligned} \label{VehicleConstraints}%
\end{equation}
where $v_{max}>0$ and $v_{min}\geq0$ denote the maximum and minimum velocity
allowed in the CZ, while $u_{i,min}<0$ and $u_{i,max}>0$ denote the minimum
and maximum control input for each CAV $i$, respectively.

Similar to prior work, we use a quadratic function for $\mathcal{C}(u_{i}(t))$
in (\ref{eqn:obj}) and thus minimize energy consumption by minimizing the
control input effort $\frac{1}{2}u_{i}^{2}(t)$. By normalizing travel time and
$\frac{1}{2}u_{i}^{2}(t)$, and using $\alpha\in\lbrack0,1)$, we construct a
convex combination as follows:
\begin{equation}
\begin{aligned}\min_{u_{i}(t),t_i^{m}} \int_{t_i^0}^{t_i^{m}}\left(\alpha + \frac{(1-\alpha)\frac{1}{2}u_i^2(t)}{\frac{1}{2}\max \{u_{max}^2, u_{min}^2\}}\right)dt \end{aligned}.
\label{eqn:energyobja}%
\end{equation}
\noindent If $\alpha=1$, the problem (\ref{eqn:energyobja}) is equivalent to a minimum
travel time problem; if $\alpha=0$, it becomes a minimum energy consumption problem.

By defining $\beta\equiv\frac{\alpha\max\{u_{\max}^{2},u_{\min}^{2}%
\}}{2(1-\alpha)}$ (assuming $\alpha<1$) and multiplying (\ref{eqn:energyobja})
by the constant $\frac{\beta}{\alpha}$, we have:
\begin{equation}
\min_{u_{i}(t),t_{i}^{m}}\beta(t_{i}^{m}-t_{i}^{0})+\int_{t_{i}^{0}}%
^{t_{i}^{m}}\frac{1}{2}u_{i}^{2}(t)dt, \label{eqn:energyobj}%
\end{equation}
\noindent where $\beta\geq0$ is a weight factor that can be adjusted through $\alpha
\in\lbrack0,1)$ to penalize travel time relative to the energy cost. Then, the
optimization problem can be stated as:

\begin{problem}
\label{prob:merg} For each CAV $i$ governed by dynamics (\ref{VehicleDynamics}%
) ignoring noise terms, determine a control law such that (\ref{eqn:energyobj}%
) is minimized subject to (\ref{VehicleDynamics}), (\ref{Safety}),
(\ref{SafeMerging}), (\ref{VehicleConstraints}), given $t_{i}^{0}$ and the
initial and final conditions $x_{i}(t_{i}^{0})=0$, $v_{i}(t_{i}^{0})$,
$x_{i}(t_{i}^{m})$.
\end{problem}

\section{Multi-lane Intersection Problem Solution}

\label{sec:solution}

The solution of \emph{Problem} \ref{prob:merg} can be obtained as described in
\cite{xiao2019decentralized} where a single MP is involved in a two-road
single-lane merging configuration where the value of $j$ in (\ref{SafeMerging}%
) is immediately known. The difficulty here is that there may be more than one
CAV $j$ defining lateral safety constraints for any $i\in S(t)$ and
determining the value(s) of $j$ is challenging since there are eight lanes and
three possible actions at intersections as shown in Fig. \ref{fig:inter}.
Obviously, this can become even harder as more lanes are added or asymmetrical
intersections are considered. Therefore, we propose a general MP-based
approach which involves two steps. The first step addresses the following two
issues: $(i)$ A strategy for determining \textquotedblleft
floating\textquotedblright\ MPs due to CAVs possibly changing lanes within the
CZ, $(ii)$ A strategy for determining all lateral safety constraints, hence
the values of $j$ in (\ref{SafeMerging}). Once these issues are addressed in
the remainder of this section, \emph{Problem} \ref{prob:merg} is well-defined.
The second step consists of solving \emph{Problem} \ref{prob:merg} and
developing the proposed joint optimal control and barrier function controller
in the next section. The overall process is outlined in \textbf{Algorithm 1}
which is implemented in time-driven manner by replanning the control inputs of
all CAVs every $T$ seconds.

\begin{algorithm}
\caption{A MP-based Algorithm for Multi-lane intersection problems}
\begin{algorithmic}[1]
\STATE Initialize an empty queue table $S(t)$.
\FOR{every $T$ seconds}
\IF{a new vehicle enters the CZ}
\STATE Determine a passing order for all CAVs according to the FIFO rule or other resequencing methods, e.g., the DR scheme.
\STATE Plan an \emph{unconstrained} optimal control trajectory for the new CAV.
\IF{the new CAV needs to change lanes}
\STATE Use the lane-changing MP determination strategy (Section III.A) to determine the lane-changing location and time for the new CAV
\ENDIF
\STATE Add the information of the new CAV into $S(t)$.
\ENDIF
\FOR{each CAV in $S(t)$}
\STATE Use the lateral safety constraint determination strategy (Section III.B) to determine which constraints it needs to meet.
\STATE Use the joint optimal control and barrier function controller (Section IV) to obtain control inputs for it.
\IF{this CAV has left the intersection}
\STATE Remove the information of this CAV from $S(t)$.
\ENDIF
\ENDFOR
\ENDFOR
\end{algorithmic}
\end{algorithm}

\subsection{Determination of Lane-changing MPs}

\label{sec:lane} When a new CAV $i\in S(t)$ arrives at the origins
$O_{2},O_{4},O_{6},O_{8}$ (or $O_{1},O_{3},O_{5},O_{7}$) and must turn left
(or right), it has to change lanes before getting close to the intersection.
Therefore, CAV $i$ must determine the location of the variable (floating) MP
$M_{i,k},$ $k\in\{1,2,\cdots,8\}$.

There are three important observations to make:

$(i)$ The \emph{unconstrained} optimal control for such $i$ is independent of
the location of $M_{i,k},$ $k\in\{1,2,\cdots,8\}$ since we have assumed that
lane-changing will only induce a fixed extra length $l$ regrdless of where it occurs.

$(ii)$ The optimal control solution under the lateral safety constraint is
better (i.e., lower cost in (\ref{eqn:energyobj})) than one which includes an
active rear-end safety constrained arc in its optimal trajectory. This is
because the former applies only to a single time instant $t_{i}^{k}$ whereas
the latter requires the constraint (\ref{Safety}) to be satisfied over all
$t\in\lbrack t_{i}^{0},t_{i}^{k}]$. It follows that any MP $M_{i,k}$ should be
as close as possible to the intersection (i.e., $L_{i,k}$ should be as large
as possible, and its maximum value is $L_{2}+L_{3}$), since the lateral safety
constraint after $M_{i,k}$ will become a rear-end safety constraint with
respect to some $j$ in the adjacent lane.

$(iii)$ In addition, CAV $i$ may also be constrained by its physically
preceding CAV $i_{p}$ (if one exists) in the same lane as $i$. In this case,
CAV $i$ needs to consider both the rear-end safety constraint with $i_{p}$ and
the lateral safety constraint with some $j\neq i$. Thus, the solution is more
constrained (hence, more sub-optimal) if $i$ stays in the current lane after
the rear-end safety constraint due to $i_{p}$ becomes active. We conclude that
in this case CAV $i$ should change its lane to the left (right) lane as late
as possible, i.e., just as the rear-end safety constraint with $i_{p}$ first
becomes active, i.e., $L_{i,k}$ is determined by
\begin{equation}
L_{i,k}=x_{i}^{\ast}(t_{i}^{a}) \label{Li1}%
\end{equation}
where $x_{i}^{\ast}(t)$ denotes the unconstrained optimal trajectory of CAV
$i$ (as determined in Sec. \ref{sec:OCBF}), and $t_{i}^{a}\geq t_{i}^{0}$ is
the time instant when the rear-end safety constraint between $i$ and $i_{p}$
first becomes active; if this constraint never becomes active, then
$L_{i,k}=L_{2}+L_{3}$. The value of $t_{i}^{a}$ is determined from
(\ref{Safety}) by
\begin{equation}
x_{i_{p}}^{\ast}(t_{i}^{a})-x_{i}^{\ast}(t_{i}^{a})=\varphi v_{i}^{\ast}%
(t_{i}^{a})+\delta, \label{tia}%
\end{equation}
where $x_{i_{p}}^{\ast}(t)$ is the \emph{unconstrained} optimal position of
CAV $i_{p}$ and $v_{i}^{\ast}(t)$ is the \emph{unconstrained} optimal speed of
CAV $i$. If, however, CAV $i_{p}$'s optimal trajectory itself happened to
include a constrained arc, then (\ref{tia}) is only an upper bound of
$t_{i}^{a}$.

In summary, it follows from $(i)-(iii)$ above that if CAV $i$ never encounters
a point in its current lane where its rear-end safety constraint becomes
active, we set $L_{i,k}=L_{2}+L_{3}$, otherwise, $L_{i,k}$ is determined
through (\ref{Li1})-(\ref{tia}).

We note that the distances from the origins $O_{1},\dots,O_{8}$ to MPs are not
all the same, and the CAVs that make a lane change will induce an extra $l$
distance. Therefore, we need to perform a coordinate transformation for those
CAVs that are in different lanes and will meet at the same MP $M_{k},$
$k\in\{1,\dots,32\}$. In other words, when $i\in S(t)$ obtains information for
$j\in S(t)$ from the FIFO queue table to account for the lateral safety
constraint at an MP $M_{k}$, the position information $x_{j}(t)$ is
transformed into $x_{j}^{\prime}(t)$ through
\begin{equation}
x_{j}^{\prime}(t):=\left\{
\begin{array}
[c]{rcl}%
x_{j}(t)+L_{i,k}-L_{j,k}+l, & \mbox{if only $i$ changes lane} & \\
x_{j}(t)+L_{i,k}-L_{j,k}-l, & \mbox{if only $j$ changes lane}, & \\
x_{j}(t)+L_{i,k}-L_{j,k}, & \mbox{otherwise}. &
\end{array}
\right.  \label{eqn:trans}%
\end{equation}
where $L_{i,k},L_{j,k}$ denote the distances of the MPs $M_{k}$ from the
origins of CAVs $i$ and $j$, respectively. Note that the coordinate
transformation (\ref{eqn:trans}) only applies to CAV $i$ obtaining information
on $j$ from $S(t)$, and does not involve any action by the coordinator.

\subsection{Determination of Lateral Safety Constraints}

\label{sec:merg_stra}

We begin with the observation (by simple inspection of Fig. 1) that CAVs can
be classified into two categories, depending on the lane that a CAV arrives
at, as follows:

\begin{enumerate}
\item The CAVs arriving at lanes $l_{1},l_{3},l_{5},l_{7}$ will pass

\begin{itemize}
\item two MPs if they choose to turn right (including the floating MP
$M_{i,k},$ $k\in\{2,4,6,8\}$);

\item four MPs it they turn left;

\item five MPs if they go straight.
\end{itemize}

\item The CAVs arriving at lanes $l_{2},l_{4},l_{6},l_{8}$ will pass

\begin{itemize}
\item only one MP if they choose to turn right;

\item five MPs if they turn left (including the floating MP $M_{i,k},$
$k\in\{1,3,5,7\}$) or if they go straight.
\end{itemize}
\end{enumerate}

\begin{figure}[ptbh]
\vspace{-0mm} \centering
\includegraphics[scale=0.20]{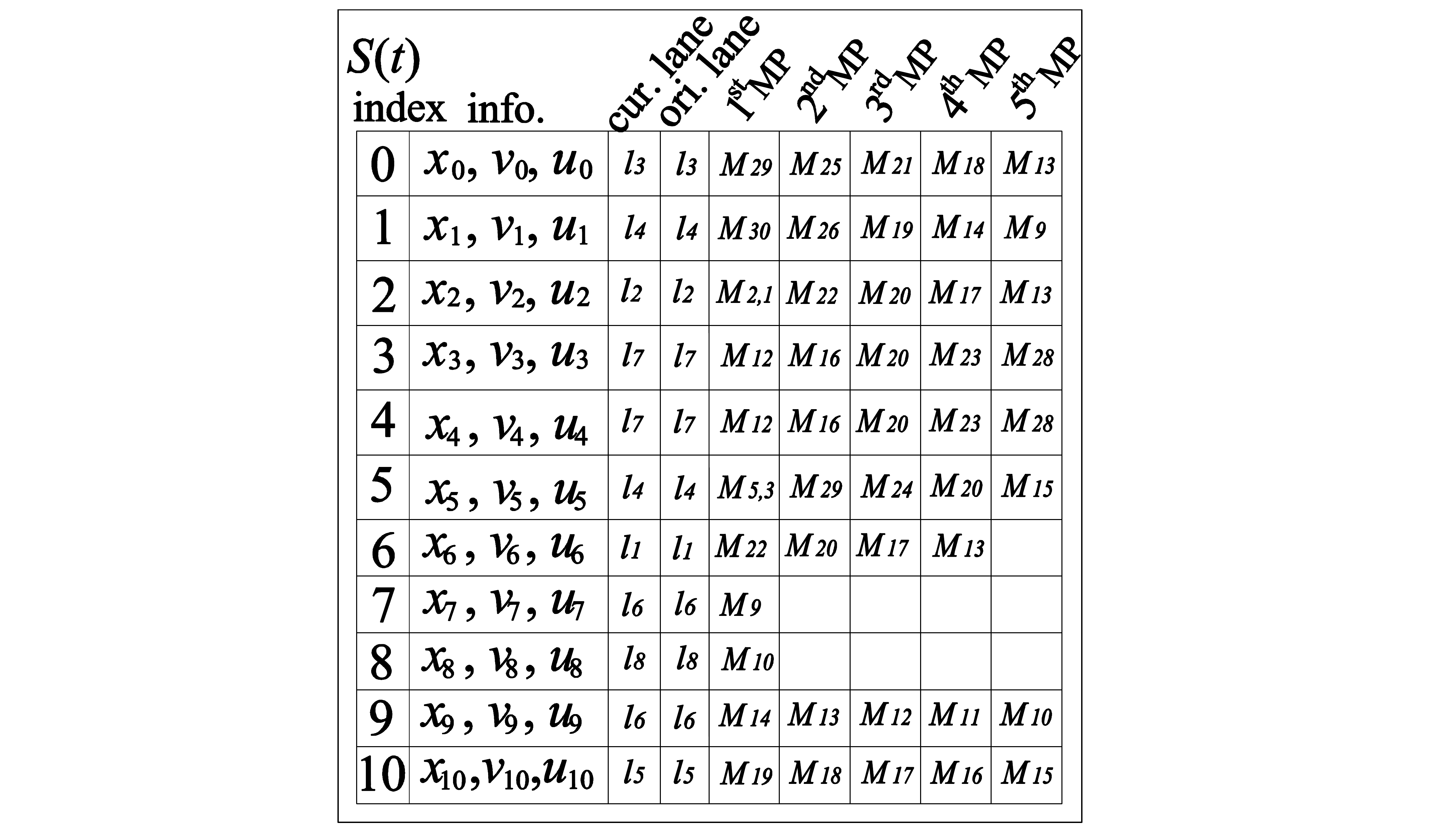} \vspace{-0mm} \vspace{-0mm}\caption{The
extended coordinator queue table.}%
\label{fig:queue}%
\end{figure}

Clearly, the maximum number of MPs a CAV may pass is 5. Since all such MPs are
determined upon arrival at the CZ, we augment the queue table in Fig.
\ref{fig:inter} by adding the original lane and the MP information for each
CAV as shown in Fig. \ref{fig:queue} for a snapshot of Fig. \ref{fig:inter}. The current and original lanes are shown
in the third and fourth column, respectively. The original lane is fixed,
while the current lane may vary dynamically: in \textbf{Algorithm 1}, the
state of all CAVs in the queue is updated if any of them has changed lanes.
The remaining five columns show all MPs a CAV will pass through in order. For
example, left-turning CAV $2$ in Fig. \ref{fig:inter} passes through five MPs
$M_{2,1},M_{22},M_{20},M_{17}$, and $M_{13}$ sequentially, where we label the
$1$st MP as $M_{2,1}$ and so forth.

When a new CAV enters the CZ, it first determines whether it will change lanes
(as described in Section III.A) and identifies all MPs that it must pass. At
this point, it looks up the extended queue table $S(t)$ (an example is shown
in Fig. \ref{fig:queue}) which already contains all prior CAV states and MP
information. First, from the current lane column, CAV $i$ can determine its
current physically immediately preceding CAV $i_{p}$ if one exists. Next,
since the passing priority has been determined by the sequencing method
selected (FIFO or otherwise), CAVs need to yield to other CAVs that rank
higher in the queue $S(t)$. In addition, for any MP CAV $i$ will pass through,
it only needs to yield to the closest CAV that has higher priority than it,
and this priority is determined by the order of $S(t)$. For instance, CAVs
$3$, $4$, and $5$ will all pass through $M_{20}$, as shown in Fig.
\ref{fig:queue}. For the MP $M_{20}$, CAV $5$ only needs to meet the lateral
safety constraint with CAV $4$, while the constraint with CAV $3$ will be
automatically met since CAV $4$ yields to CAV $3$. Similarly, we can find
indices of CAVs for other MPs crossed by CAV $5$. Since CAV $5$ passes through
five MPs, we define an index set $\Omega_{5}$ for CAV $5$ which has at most 5
elements. In this example, CAV $5$ only conflicts with CAV $0$ at $M_{29}$
besides $M_{20}$, so that $\Omega_{5}=\{0,4\}$.

\begin{algorithm}
\caption{Search Algorithm for Conflict CAVs}
\begin{algorithmic}[1]
\REQUIRE The extended coordinator queue table $S(t)$, CAV $i$
\ENSURE The index of conflict CAVs for CAV $i$
\STATE Initialize an empty index set $\Omega_i$.
\STATE Initialize a set $\Theta_i$ including all MPs CAV $i$ will pass through
\STATE Find the position $k$ of CAV $i$ in the $S(t)$.
\FOR{$j = k-1 : -1 : 1$}
\IF{the $j$th CAV in $S(t)$ passes at least one MP in $\Theta_i$}
\STATE Add the index of this CAV into the set $\Omega_i$.
\STATE Remove the same MPs from $\Theta_i$.
\ENDIF
\IF {$\Theta_i$ is empty}
\STATE break;
\ENDIF
\ENDFOR
\RETURN $\Omega_i$
\end{algorithmic}
\end{algorithm}

Therefore, it remains to use the information in $S(t)$ in a systematic way so
as to determine all the indices of those CAVs that CAV $i$ needs to yield to;
these define each index $j$ in (\ref{SafeMerging}) constituting all lateral
safety constraints that CAV $i$ needs to satisfy. This is accomplished by a
search algorithm (\textbf{Algorithm 2}) based on the following process. CAV
$i$ compares its original lane and MP information to that of every CAV in the
table starting with the last row and moving up. The process terminates the
first time that any one of the following three conditions is satisfied at some
row $j<i$:

\begin{enumerate}
\item All MP information of CAV $i$ matches row $j$ and $\Omega_{i}$ is empty.

\item Every MP for CAV $i$ has been matched to some row $j$.

\item All prior rows $j<i$ have been looked up.
\end{enumerate}

These three conditions are examined in order:

\textbf{Condition (1)}: If this is satisfied, there are no conflicting MPs for
CAV $i$ and this implies that CAV $i_{p}$ is the physically immediately
preceding CAV all the way through the CZ. Thus, CAV $i$ only has to satisfy
the safety constraint (\ref{Safety}) with respect to $i_{p}$, i.e., it just
follows CAV $i_{p}$. For example, $i=4$ and $i_{p}=3$ in Fig. \ref{fig:inter}
(and Fig. \ref{fig:queue}).

\textbf{Condition (2)}: If this holds, then CAV $i$ has to satisfy several
lateral safety constraints with one or more CAV $j\in\Omega_{i}$. Moreover, it
also has to satisfy the rear-end safety constraint (\ref{Safety}) with CAV
$i_{p}$, where $i_{p}$ is determined by the first matched row in the current
lane column of Fig. \ref{fig:queue}. For example, $i=10$, $j=0,1,4,5$, and $6$
in Fig. \ref{fig:inter} (and Fig. \ref{fig:queue}).

\textbf{Condition (3):} There are two cases. First, if the index set
$\Omega_{i}$ is empty, then CAV $i$ does not have to satisfy any lateral
safety constraint; for example, $i=7$ in Fig. \ref{fig:inter}. Otherwise, it
needs to yield to all CAVs in $\Omega_{i}$; for example, $i=2$, $j=0$
in Fig. \ref{fig:inter}.

We observe that \textbf{Algorithm 2} can be implemented for all CAVs in an
event-driven way (since $S(t)$ needs to be updated only when an event that
changes its state occurs). The triggering events are: $(i)$ a CAV entering the
CZ, $(ii)$ a CAV departing the CZ, $(iii)$ a CAV completing a lane change at a floating MP, and
$(iv)$ a CAV overtaking event (a lane change event at a fixed MP). This last event may occur when a CAV merges into another 
lane at a MP through which it leaves the CZ. In particular, consider three CAVs $i,j$, and $k$ such that $k > j > i$, and CAV $j$ merges into the same lane as $i$ and $k$. Then, CAV $k$ looks at the first row above it where there is a CAV with the same lane; that’s now CAV $j$. However, $i$ is physically ahead of $k$. Thus, we need to re-order the queue according to the incremental
position order, so that a CAV following $i$ (CAV $k$) can properly
identify its physically preceding CAV. For example, consider $i=7$, $j=8$, and
$k=9$ in Fig. \ref{fig:inter}, and suppose that CAV 7 turns right, CAV 8 turns right, and
CAV 9 goes straight. Their order in $S(t)$ is 7, 8, 9. CAV 8
can overtake CAV 7, and its current lane will become $l_{6}$ when it passes all
MPs. Since CAV 7 and CAV 9 also are in $l_{6}$, CAV 9 will mistake CAV 8 as
its new preceding CAV after the current lane of CAV 8 is updated. However, in
reality CAV 7 is still the preceding CAV of CAV 9, hence CAV 9 may
accidentally collide with CAV 7. To avoid this problem, we need to re-order
the queue according to the position information when this event occurs, i.e.,
making CAV 8 have higher priority than CAV 7 in the queue. Alternatively, this
problem may be resolved by simply neglecting CAVs that have passed all MPs
when searching for the correct identity of the CAV that precedes $i$.

\section{Joint Optimal and Control Barrier Function Controller}

\label{sec:OCBF}Once a newly arriving CAV $i\in S(t)$ has determined all the
lateral safety constraints (\ref{SafeMerging}) it has to satisfy, it can solve
problem (\ref{eqn:energyobj}) subject to these constraints along with the
rear-end safety constraint (\ref{Safety}) and the state limitations
(\ref{VehicleConstraints}). Obtaining a solution to this constrained optimal
control problem is computationally intensive, as shown in the single-lane
merging problem \cite{xiao2019decentralized}, and this complexity is obviously
higher for our multi-lane intersection problem since there are more lateral
safety constraints. Therefore, in this section, we proceed in two steps: $(i)$
We solve the \emph{unconstrained} version of problem (\ref{eqn:energyobj})
which can be done with minimal computational effort, and $(ii)$ We optimally
track the unconstrained problem solution while using CBFs to account for all
constraints and guarantee they are never violated (as well as Control Lyapunov
Functions (CLFs) to better track the unconstrained optimal states). Since this
controller $u_{i}(t)$ for CAV $i$ combines an optimal control solution with
CBFs, we refer to it as the \emph{Optimal Control and Barrier Function (OCBF)}
control. Note that all CAVs can solve Problem \ref{prob:merg} in a
decentralized way.

\subsection{Unconstrained decentralized optimal control solution}

As mentioned above, we use the CAV trajectory obtained from the unconstrained
optimal solution Problem \ref{prob:merg} as a \emph{reference} trajectory and
deal with all constraints through our OCBF controller. When all state and
safety constraints are inactive, we can obtain an analytical solution of
Problem \ref{prob:merg}. As shown in \cite{zhang2018decentralized}, the
optimal control, speed, and position trajectories are given by%

\begin{equation}
u_{i}^{\ast}(t)=a_{i}t+b_{i}\vspace{-2mm} \label{Optimal_u}%
\end{equation}%
\begin{equation}
v_{i}^{\ast}(t)=\frac{1}{2}a_{i}t^{2}+b_{i}t+c_{i}\vspace{-2mm}
\label{Optimal_v}%
\end{equation}%
\begin{equation}
x_{i}^{\ast}(t)=\frac{1}{6}a_{i}t^{3}+\frac{1}{2}b_{i}t^{2}+c_{i}t+d_{i}
\label{Optimal_x}%
\end{equation}
where $a_{i}$, $b_{i}$, $c_{i}$ and $d_{i}$ are integration constants that can
be solved along with{\ $t_{i}^{m}$} by the following five algebraic equations
(details given in \cite{zhang2018decentralized}):
\begin{equation}
\begin{aligned} &\frac{1}{2}a_i\cdot(t_i^0)^2 + b_it_i^0 + c_i = v_i^0,\\ &\frac{1}{6}a_i\cdot(t_i^0)^3 + \frac{1}{2}b_i\cdot(t_i^0)^2 + c_it_i^0+d_i = 0,\\ &\frac{1}{6}a_i\cdot(t_i^{m})^3 + \frac{1}{2}b_i\cdot(t_i^{m})^2 + c_it_i^{m}+d_i = L_k,\\ &a_it_i^{m} + b_i = 0,\\ &\beta - \frac{1}{2}b_i^2+a_ic_i = 0. \end{aligned} \label{OptimalSolInA}%
\end{equation}
where the third equation is the terminal condition for the total distance
traveled on a lane. This solution is computationally very efficient to obtain.
We then use this unconstrained optimal control solution as a reference to be
tracked by a controller which uses CBFs to account for all the constraints
(\ref{Safety}), (\ref{SafeMerging}), and (\ref{VehicleConstraints}) and
guarantee they are not violated. We review next how to use CBFs to map all
these constraints from the state $x_{i}(t)$ to the control input $u_{i}(t)$.

\subsection{CBF controller}

The OCBF controller aims to track the unconstrained optimal control solution
(\ref{Optimal_u})-(\ref{Optimal_x}) while satisfying all constraints
(\ref{Safety}), (\ref{VehicleConstraints}) and (\ref{SafeMergingc}). To
accomplish this, first let $\bm x_{i}(t)\equiv(x_{i}(t),v_{i}(t))$. Referring
to the vehicle dynamics (\ref{VehicleDynamics}), let $f(\bm x_{i}%
(t))=[x_{i}(t),0]^{T}$ and $g(\bm x_{i}(t))=[0,1]^{T}$. Each of the
constraints in (\ref{Safety}), (\ref{VehicleConstraints}) and
(\ref{SafeMergingc}) can be expressed as $b_{k}(\bm x_{i}(t))\geq0$,
$k\in\{1,\cdots,n\}$ where $n$ is the number of constraints and each
$b_{k}(\bm x_{i}(t))$ is a CBF. For example, $b_{1}(\bm x_{i}(t))=z_{i,i_{p}%
}(t)-\varphi v_{i}(t)-\delta$ for the rear-end safety constraint
(\ref{Safety}). In the CBF approach, each of the continuously differentiable
\emph{state} constraints $b_{k}(\bm x_{i}(t))\geq0$ is mapped onto another
constraint on the \emph{control} input such that the satisfaction of this new
constraint implies the satisfaction of the original constraint $b_{k}(\bm
x_{i}(t))\geq0$. The forward invariance property of this method
\cite{ames2019control,Xiao2019} ensures that a control input that satisfies
the new constraint is guaranteed to also satisfy the original one. In
particular, each of these new constraints takes the form
\begin{equation}
\begin{aligned} L_fb_k(\bm x_i(t)) + L_gb_k(\bm x_i(t))u_i(t) + \gamma( b_k(\bm x_i(t))) \geq 0, \end{aligned}\label{eqn:cbf}%
\end{equation}
where $L_{f},L_{g}$ denote the Lie derivatives of $b_{k}(\bm x_{i}(t))$ along
$f$ and $g$ (defined above from the vehicle dynamics) respectively and
$\gamma(\cdot)$ denotes a class of $\mathcal{K}$ functions
\cite{khalil2002nonlinear} (typically, linear or quadratic functions). As an
alternative, a CLF \cite{ames2019control} $V(\bm x_{i}(t))$, instead of
$b_{k}(\bm x_{i}(t))$, can also be used to track (stabilize) the optimal speed
trajectory (\ref{Optimal_v}) through a CLF constraint of the form%
\begin{equation}
\begin{aligned} L_fV(\bm x_i(t)) + L_gV(\bm x_i(t))u_i(t) + \epsilon V(\bm x_i(t)) \leq e_{i}(t), \end{aligned}\label{CLF}%
\end{equation}
where $\epsilon>0$ and $e_{i}(t)$ is a relaxation variable that makes this
constraint soft. As is usually the case, we select $V(\bm x_{i}(t))=(v_{i}%
(t)-v_{ref}(t))^{2}$ where $v_{ref}(t)$ is the reference speed to be tracked
(specified below). Therefore, the OCBF controller solves the following
problem:
\begin{equation}
\small
\min_{u_{i}(t),e_{i}(t)}J_{i}(u_{i}(t),e_{i}(t))\!=\!\int_{t_{i}^{0}%
}^{t_{i}^{m}}\!\left(  \beta e_{i}^{2}(t)\!+\!\frac{1}{2}(u_{i}(t)\!-\!u_{ref}%
(t))^{2}\right)  dt,
\label{eqn:ocbf}%
\end{equation}
subject to the vehicle dynamics (\ref{VehicleDynamics}), the CBF constraints
(\ref{eqn:cbf}) and the CLF constraint (\ref{CLF}). The obvious selection for
speed and acceleration reference signals is $v_{ref}(t)=v_{i}^{\ast}(t)$,
$u_{ref}(t)=u_{i}^{\ast}(t)$ with $v_{i}^{\ast}(t)$, $u_{i}^{\ast}(t)$ given
by (\ref{Optimal_v}) and (\ref{Optimal_u}) respectively. In \cite{xiao2019decentralizedITSC},
$u_{ref}(t)=\frac{x_{i}^{\ast}(t)}{x_{i}(t)}u_{i}^{\ast}(t)$ is used to
provide the benefit of feedback provided by observing the actual CAV
trajectory $x_{i}(t)$ and automatically reducing the tracking position error;
we use only $u_{ref}(t)=u_{i}^{\ast}(t)$ in the sequel for simplicity.

The CBF conversions from the original constraints to the form
(\ref{eqn:cbf}) are straightforward. For example, using a linear function
$\gamma(\cdot)$ in (\ref{eqn:cbf}), we can directly map \textbf{Constraint 1}
into the following constraint in terms of control inputs:
\begin{equation}
\underbrace{v_{i_{p}}(t)-v_{i}(t)}_{L_{f}b(\bm x_{i}(t))}+\underbrace{-\varphi
}_{L_{g}b(\bm x_{i}(t))}u_{i}(t)+z_{i,i_{p}}(t)-\varphi v_{i}(t)-\delta
\geq0.\label{SafetyCBF1}%
\end{equation}

\noindent However, there are some points that deserve some further
clarification as follows.

\textbf{Constraint 2} (Lateral safety constraint): The lateral safety
constraints in (\ref{SafeMerging}) are specified only at time instants
$t_{i}^{k}$. However, to use CBFs as in (\ref{eqn:cbf}), they have to be
converted to \emph{continuously differentiable forms}. Thus, we use the same
technique as in \cite{Wei2019} to convert (\ref{SafeMerging}) into:
\begin{equation}
z_{i,j}(t)\geq\Phi(x_{i}(t))v_{i}(t)+\delta,\text{ }i\in S(t),\text{ }%
t\in\lbrack t_{i}^{0},t_{i}^{k}],\label{SafeMergingc}%
\end{equation}

\noindent where $j\in\Omega_{i}$ is determined through the lateral safety
constraint determination strategy (\textbf{Algorithm 2}). Recall that CAV $j$
depends on some MP $M_{k}$ and we may have several $j\in\Omega_{i}$ since CAV
$i$ may conflict with several CAVs $j$ at different MPs. The selection of
function $\Phi:\mathbb{R}\rightarrow\mathbb{R}$ is flexible as long as it is a
strictly increasing function that satisfies $\Phi(x_{i}
(t_{i}^{0}))=0$ and $\Phi(x_{i}%
(t_{i}^{k}))=\varphi$ where $t_{i}^{k}$ is the arrival time at MP $M_{k}$
corresponding to the constraint and $x_{i}(t_{i}^{k})$ is the location of MP
$M_{k}$. Thus, we see that at $t=t_{i}^{k}$ all constraints in
(\ref{SafeMergingc}) match the safe-merging constraints (\ref{SafeMerging}),
{and that at $t=t_{i}^{0}$ we have $z_{i,i_{p}}(t_{i}^{0})=\delta$}. Since the
selection of $\Phi(\cdot)$ is flexible, for simplicity, we define it to have
the linear form $\Phi(x_{i}(t))=\frac{\varphi}{x_{i}(t_{i}^{k})}x_{i}(t)$
which we can see satisfies the properties above.

\textbf{Improving the feasibility of Constraints 1 and 2:} In order to ensure
that a feasible solution always exists for these constraints, we need to take
the braking distance into consideration. CAV $i$ should stop within a minimal
safe distance when its speed $v_{i}(t)$ approaches the speed $v_{j}(t)$ for
any $j$ such that CAV $j$ is the preceding vehicle of CAV $i$ or any vehicles
that may laterally collide with CAV $i$. Thus, we use the following more
strict constraint when $v_{i}(t)\geq v_{j}(t)$:%

\begin{equation}
z_{i,j}(t) \geq\frac{\varphi\big(x_{i}(t)+\frac{1}{2}\frac{v_{i}^{2}%
(t)-v_{j}^{2}(t)}{u_{\min}}\big)v_{j}(t)}{L}+\frac{1}{2}\frac{(v_{j}%
(t)-v_{i}(t))^{2}}{u_{\min}}+\delta, \label{SafeMergingBraking}%
\end{equation}

\noindent A detailed analysis of this constraint is given in \cite{Wei2019}.

Observing that \textbf{Constraint 3} (vehicle limitations) can be directly
converted using the standard CBF method, we are now in a position where all
constraints are mapped into constraints expressed in terms of control inputs.
We refer to the resulting control $u_{i}(t)$ in (\ref{eqn:ocbf}) as the OCBF
control. The solution to (\ref{eqn:ocbf}) is obtained by discretizing the time
interval $[t_{i}^{0},t_{i}^{m}]$ with time steps of length $\Delta$ and
solving {(\ref{eqn:ocbf}) over }$[t_{i}^{0}+k\Delta,t_{i}^{0}+(k+1)\Delta]$,
$k=0,1,\ldots$, {with $u_{i}(t),$ $e_{i}(t)$ as decision variables held
constant over each such interval (see also \cite{Wei2019}). Consequently, each
such problem is a Quadratic Program (QP) since we have a quadratic cost and a
number of linear constraints on the decision variables at the beginning of
each time interval. The solution of each such problem gives an optimal control
$u_{i}^{\ast}(t_{i}^{0}+k\Delta)$}, $k=0,1,\ldots$, allowing us{\ to update
(\ref{VehicleDynamics}) in the }$k^{th}$ time interval. This process is
repeated until CAV $i$ leaves the CZ.

\subsection{The influence of noise and complicated vehicle dynamics}

Aside from the potentially long computation time, other limitations of the OC
controller include: $(i)$ It only plans the optimal trajectory once. However,
the trajectory may violate safety constraints due to noise in the vehicle
dynamics and control accuracy; $(ii)$ The OC analytical solution is limited to
simple vehicle dynamics as in (\ref{VehicleDynamics}) and becomes difficult to
obtain when more complicated vehicle dynamics are considered to better match
realistic operating conditions. For instance, in practice, we usually need to
control the input driving force of an engine instead of directly controlling
acceleration. Compared with the OC method, our OCBF approach can effectively
deal with the above problems with only slight modifications as described next.

First, due to the presence of noise, constraints may be temporarily violated,
which prevents the CBF method from satisfying the forward invariance property.
Thus, when a constraint is violated at time $t_{1}$, i.e., $b_{k}(\bm
x_{i}(t_{1}))<0$, we add a threshold to the original constraint as follows:%

\begin{equation}
\begin{aligned} L_fb_k(\bm x_i(t)) + L_gb_k(\bm x_i(t))u_i(t) \geq c_k(t), \end{aligned} \label{eqn:cbfnoise}%
\end{equation}

\noindent where $c_{k}(t)\geq0$ is a large enough value so that $b_{k}%
(x_{i}(t))$ is strictly increasing even if the system is under the worst
possible noise case. Since it is hard to directly determine the value for
$c_{k}(t)$, we add it to the objective function and have%

\begin{equation}
\small 
\min_{u_{i}(t),e_{i}(t),c(t)}\quad\int_{t_{i}^{0}}^{t_{i}^{m}%
}\!\left(  \beta e_{i}^{2}(t)\!+\!\frac{1}{2}(u_{i}(t)\!-\!u_{ref}(t))^{2} -
\eta c_{k}(t)\right)  dt, 
\label{eqn:ocbfnoise}%
\end{equation}

\noindent where $\eta$ is a weight parameter. If there are multiple
constraints that are violated at one time, we rewrite them all as
(\ref{eqn:cbfnoise}) and add all thresholds into the optimization objective.
Starting from $t_{1}$, we use the constraint (\ref{eqn:cbfnoise}) and
objective function (\ref{eqn:ocbfnoise}) to replace the original CBF
constraint and objective function, and $b_{k}(x_{i}(t))$ will be positive
again in finite time since it is increasing. When $b_{k}(x_{i}(t))$ becomes
positive again, we revert to the original CBF constraint.

Next, considering vehicle dynamics, there are numerous models which achieve
greater accuracy than the simple model (\ref{VehicleDynamics}) depending on
the situation of interest. As an example, we consider the following frequently
used nonlinear model:%

\begin{equation}
\left[
\begin{array}
[c]{c}%
\dot{x}_{i}(t)\\
\dot{v}_{i}(t)
\end{array}
\right]  =\left[
\begin{array}
[c]{c}%
v_{i}(t)\\
-\frac{1}{m_{i}}F_{r}(v_{i}(t))
\end{array}
\right]  + \left[
\begin{array}
[c]{c}%
0\\
\frac{1}{m_{i}}%
\end{array}
\right]  u_{i}(t), \label{VehicleDynamicsNonlinear}%
\end{equation}

\noindent where $m_{i}$ denotes the vehicle mass and $F_{r}(v_{i}(t))$ is the
resistance force that is normally expressed as%

\begin{equation}
F_{r}(v_{i}(t)) = \alpha_{0} sgn(v_{i}(t))+\alpha_{1} v_{i}(t)+\alpha_{2}%
v_{i}^{2}(t),
\end{equation}

\noindent where $\alpha_{0}>0$, $\alpha_{1}>0$, and $\alpha_{2}>0$ are
parameters determined empirically, and $sgn(\cdot)$ is the signum function. It
is clear that due to the nonlinearity in these vehicle dynamics, it is
unrealistic to expect an analytical solution for it. However, in our proposed
OCBF method, we only need to derive the Lie derivative along these new
dynamics and solve the corresponding QP based on these new CBF constraints.
For instance, it is easy to see that for these new dynamics, the CBF
constraint (\ref{SafetyCBF1}) becomes%
\begin{equation}
\begin{aligned}
&\underbrace{v_{i_{p}}(t)-v_{i}(t)+\frac{\varphi F_{r}(v_{i}(t))}{m_{i}}%
}_{L_{f}b(\bm x_{i}(t))}+\underbrace{-\frac{\varphi}{m_{i}}}_{L_{g}%
b(\bm x_{i}(t))}u_{i}(t) \\
& +z_{i,i_{p}}(t)-\varphi v_{i}(t)- \delta
\geq0.\label{HOCBFcontrolnew}%
\end{aligned}
\end{equation}

Thus, our method can be easily extended to more complicated vehicle dynamics
dictated by any application of interest.

\bigskip

\section{SIMULATION RESULTS}

To validate the effectiveness of the proposed OCBF method, we compare it to a
state-of-the-art method in \cite{zhang2019decentralized} where CAVs calculate
the fastest arrival time to the conflict zone first when they enter the CZ and
then derive an energy-time-optimal trajectory. This uses the same objective
function (\ref{eqn:energyobj}) as our method. The main differences are: 1) it
considers the merging (conflict) zone as a whole and imposes the conservative
requirement that any two vehicles that have potential conflict cannot be in
the conflict zone at the same time; 2) when it plans an energy-time-optimal
trajectory for a new incoming vehicle, it takes all safety constraints into
account and thus is time-consuming; 3) the rear-end safety constraints used in
\cite{zhang2019decentralized} only depends on distance, i.e., $\varphi=0$ and
$\delta>0$ in (\ref{Safety}). Thus, in order to carry out a fair comparison
with it, we adopt the same form of rear-end safety constraints, that is,%
\begin{equation}
z_{i,i_{p}}(t)\geq\delta,\text{ \ }\forall t\in\lbrack t_{i}^{0},t_{i}%
^{m}].\label{SafetyNew}%
\end{equation}
\noindent A complication caused by this choice is that after using the
standard CBF method (simply substituting $\varphi=0$ into (\ref{SafetyCBF1})),
the control input should satisfy%
\begin{equation}
\underbrace{v_{i_{p}}(t)-v_{i}(t)}_{L_{f}b(\bm x_{i}(t))}+\underbrace{0}%
_{L_{g}b(\bm x_{i}(t))}u_{i}(t)+z_{i,i_{p}}(t)-\delta\geq
0,\label{SafetyNewCBF}%
\end{equation}
\noindent which violates the condition $L_{g}b(\bm x_{i}(t))\neq0$. This is
because we cannot obtain a relationship involving the control input $u_{i}(t)$
from the first-order derivative of the constraint (\ref{SafetyNew}). This
problem was overcome in \cite{Xiao2019} by using a high order CBF
(HOCBF) of relative degree 2 for system (\ref{VehicleDynamics}). In
particular, letting $b_{k}(\bm x_{i}(t))=z_{i,i_{p}}(t)-\delta$ and
considering all class $\mathcal{K}$ functions to be linear functions, we
define
\begin{equation}%
\begin{split}
&  \psi_{1}(\bm x_{i}(t))=\dot{b}(\bm x_{i}(t))+pb(\bm x_{i}(t),\\
&  \psi_{2}(\bm x_{i}(t))=\dot{\psi_{1}}(\bm x_{i}(t))+p\psi_{1}(\bm
x_{i}(t)).
\end{split}
\label{eq:HOCBF}%
\end{equation}
\noindent where $p$ is a (tunable) penalty coefficient. Combining the vehicle
dynamics (\ref{VehicleDynamics}) with (\ref{eq:HOCBF}), any control input
should satisfy%
\begin{equation}
\underbrace{0}_{L_{f}^{2}b(\bm x_{i}(t))}+\underbrace{-1}_{L_{g}L_{f}b(\bm
x_{i}(t))}u_{i}(t)+2p\dot{b}(\bm x_{i}(t))+p^{2}b(\bm x_{i}(t))\geq
0.\label{HOCBFcontrol}%
\end{equation}
\noindent Thus, in the following simulation experiments, we set $\varphi$ for
\textbf{Constraint 1} and \textbf{Constraint 2} to be $\varphi=0$ and
$\varphi=1.8s$, respectively.

Our simulation experiments are organized as follows. First, in Section V.A we
consider an intersection with a single lane in each direction which only
allows going straight. Our purpose here is to show that using MPs instead of 
an entire arbitrarily defined conflict zone can effectively reduce the 
conservatism of the latter. Then, in Section V.B, we allow turns so as to
analyze the influence of different behaviors (going straight, turning left,
and turning right) on the performance of the methods compared. Next, Section
V.C is intended to validate the effectiveness of our OCBF method for
intersections with two lanes and include possible lane-changing behaviors. In
Section V.D, we extend our method to combine it with the DR method and show
that the performance of the OCBF+DR method is better than the OCBF+FIFO method
for asymmetrical intersections. Finally, Section V.E demonstrates that our
method can effectively deal with complicated vehicle dynamics and noise.

The baseline for our simulation results uses SUMO, a microscopic traffic
simulation software package. Then, we use our OCBF controller and the
controller proposed in \cite{zhang2019decentralized} to control CAVs for
intersection scenarios with the same vehicle arrival patterns as SUMO. The
parameter settings (see Fig. \ref{fig:inter}) are as follows: $L1=300m$,
$L2=50m$, $L_{3}=200m$, $l=0.9378m$, $w=3.5m$, $r=4m$, $\delta=10m$, $v_{\max
}=15m/s$, $v_{\min}=0m/s$, $u_{\max}=3m/s^{2}$, and $u_{\min}=-3m/s^{2}$.

The energy model we use in the objective function is an approximate one. The
$\frac{1}{2}u^{2}$ metric treats acceleration and deceleration the same and
does not account for speed as part of energy. This metric is viewed as a
simple surrogate function for energy or simply as a measure of how much the
solution deviates from the ideal constant-speed trajectory. In contrast, the
following energy model \cite{kamal2012model} captures fuel consumption in
detail and provides another measure of performance:%

\begin{equation}%
\begin{split}
&  F_{i} = \int_{0}^{a_{i}} f_{V,i}(t) dt,\\
&  f_{V,i}(t) = f_{cruise,i}(t) + f_{accel,i}(t),\\
&  f_{cruise,i}(t) = b_{0} + b_{1}v_{i}(t) + b_{2}v_{i}^{2}(t) + b_{3}%
v_{i}^{3}(t),\\
&  f_{accel,i}(t) = u(t)(c_{0} + c_{1}v_{i}(t) + c_{2}v_{i}^{2}(t)),
\end{split}
\label{eq:fuel}%
\end{equation}

\noindent where $f_{cruise,i}(t)$ denotes the fuel consumed per second when
CAV $i$ drives at a steady velocity $v_{i}(t)$, and $f_{accel,i}(t)$ is the
additional fuel consumed due to the presence of positive acceleration. If
$u(t)\leq0$, then $f_{accel,i}(t)$ will be 0 since the engine is rotated by
the kinetic energy of the CAV in this case. $b_{0},b_{1},b_{2},b_{3}%
,c_{0},c_{1}$, and $c_{2}$ are seven model parameters, and here we use the
same parameters as in \cite{kamal2012model}, which are obtained through
curve-fitting for data from a typical vehicle.

\subsection{MPs versus conflict zone}

In this experiment, we only allow CAVs to go straight in order to investigate
the relative performance of the MP-based method (our OCBF controller) and the
merging (conflict) zone-based method (OC controller) \cite{zhang2019decentralized}. We
set the arrival rates at all lanes the same, i.e., 270 veh/h/lane. The
comparison results are shown in Table I.

\begin{table}[ptbh]
\centering
\begin{threeparttable}
\caption{The comparison results for a single-lane intersection disallowing turns}
\begin{tabular}{cccccc}
\toprule
$\beta$  & Methods & Energy & Travel time (s) & Fuel (mL) & Ave. obj.$^1$ \\
\midrule
\multirow{3}[2]{*}{0.1} & SUMO  & 23.1788 & 28.3810 & 30.3597 & 26.0169 \\
& OC    & 0.1498 & 28.3884 & 14.6266 & 2.9886 \\
& OCBF  & 0.9501 & 25.0863 & 18.3088 & 3.4587 \\
\midrule
\multirow{3}[2]{*}{0.5} & SUMO  & 23.1788 & 28.3810 & 30.3597 & 37.3693 \\
& OC    & 0.6515 & 26.0315 & 17.1585 & 13.6673 \\
& OCBF  & 2.1708 & 22.6623 & 18.9396 & 13.5020 \\
\midrule
\multirow{3}[2]{*}{1} & SUMO  & 23.1788 & 28.3810 & 30.3597 & 51.5598 \\
& OC    & 0.8782 & 25.6961 & 17.1555 & 26.5743 \\
& OCBF  & 2.9106 & 22.2617 & 18.9589 & 25.1723 \\
\midrule
\multirow{3}[2]{*}{2} & SUMO  & 23.1788 & 28.3810 & 30.3597 & 79.9408 \\
& OC    & 1.1869 & 25.4834 & 17.1658 & 52.1537 \\
& OCBF  & 3.9157 & 21.9139 & 18.9852 & 47.7435 \\
\bottomrule
\end{tabular}%
\begin{tablenotes}
\footnotesize
\item[1] Ave. obj. = $\beta \times$ Travel time + Energy.
\end{tablenotes}
\end{threeparttable}
\label{tab:table1}\end{table}

It is clear that both controllers significantly outperform the results
obtained from the SUMO controller which employs a standard car following
model. The OC controller is the energy-optimal since it has considered all
safety constraints at the initial time for each CAV. Then, CAVs strictly
follow the planned trajectory assuming that noise is absent. However, for our
controller, we only use an \emph{unconstrained} reference trajectory and
employ CBFs to account for the fact that the reference trajectory may violate
the safety constraints: the OCBF controller in each CAV continuously updates
its control inputs according to the latest states of other CAVs. As a result,
its energy consumption is larger than that of the OC controller, although
still small and much less than under the SUMO car following controller.

In terms of travel time, we find that the travel time of the OCBF controller
is better than that of the OC controller. This is because the safety
requirements in the OC controller are too strict so that a CAV $i$ must wait
until a CAV $j\neq i$ that conflicts with it leaves the conflict zone.
Instead, the OCBF controller using MPs allows us to relax a merging constraint
and still ensure safety by requiring that one CAV only can arrive at the same
MP $\varphi$ seconds after the other vehicle leaves. Since our method reduces
conservatism, it shows significant improvement in travel time when compared
with the OC controller in \cite{zhang2019decentralized}.

In addition, we can adjust the parameter $\beta$ to emphasize the relative
importance of one objective (energy or time) relative to the other. If we are
more concerned about energy consumption, we can use a smaller $\beta$;
otherwise, we can use a larger $\beta$ to emphasize travel time reduction.
Thus, when $\beta$ is relatively large, the average objective of the OCBF
controller is better than that of the OC controller since it is more efficient
with respect to travel time.

Another interesting observation is that even though the relationship between
the accurate fuel consumption model and the estimated energy is complicated,
we see that a larger estimated energy consumption usually corresponds to
larger fuel consumption. Thus, it is reasonable to optimize energy consumption
through a simple model, e.g., $\frac{1}{2}u^{2}$, which also significantly
reduces the computational complexity caused by the accurate energy model.

\subsection{The influence of turns}

In this experiment, we allow turns at the intersection assuming that the
behavior of a CAV when it enters the CZ is known. We have conducted four
groups of simulations as shown in Table \ref{tab:table2}. In the first group,
all CAVs choose their behavior with the same probability, i.e., $\frac{1}{3}$
going straight, $\frac{1}{3}$ turning left, and $\frac{1}{3}$ turning right.
In the second group, 80\% of CAVs turn left while 10\% CAVs go straight and
10\% CAVs turn right. In the third group, 80\% CAVs turn right while 10\% CAVs
go straight and 10\% CAVs turn left. In the fourth group, 80\% CAVs go
straight while 10\% CAVs turn left and 10\% CAVs turn right. We set $\beta=1$
in all results shown in Table \ref{tab:table2}.

\begin{table}[ptbh]
\caption{The influence of turns on different controllers}%
\label{tab:table2}
\centering
\begin{tabular}
[c]{cccccc}%
\toprule Groups & Methods & Energy & Travel time & Fuel & Ave. obj.\\
\midrule \multirow{3}[2]{*}{1} & SUMO & 23.7011 & 28.2503 & 28.3357 &
51.9514\\
& OC & 1.2957 & 22.4667 & 18.8433 & 23.7624\\
& OCBF & 2.6193 & 21.8514 & 18.9208 & 24.4707\\
\midrule \multirow{3}[2]{*}{2} & SUMO & 26.5612 & 31.3822 & 29.0524 &
57.9434\\
& OC & 1.0666 & 24.2179 & 18.0072 & 25.2845\\
& OCBF & 2.4537 & 21.8707 & 18.8704 & 24.3244\\
\midrule \multirow{3}[2]{*}{3} & SUMO & 19.9066 & 24.2937 & 25.6778 &
44.2003\\
& OC & 1.3775 & 22.0706 & 18.8803 & 23.4481\\
& OCBF & 2.3623 & 21.4874 & 18.8306 & 23.8497\\
\midrule \multirow{3}[2]{*}{4} & SUMO & 21.7450 & 26.6148 & 29.4484 &
48.3598\\
& OC & 1.2602 & 22.7417 & 18.7339 & 24.0019\\
& OCBF & 2.5884 & 22.0114 & 18.9230 & 24.5998\\
\bottomrule &  &  &  &  &
\end{tabular}
\end{table}

First, we can draw the same conclusion as in Table I that the OC controller
is energy-optimal and the OCBF controller achieves the smallest travel time
since it reduces conservatism. Next, we also observe that when we increase the
ratio of left-turning vehicles, the average travel times under all controllers
increase; when we increase the ratio of right-turning vehicles, the average
travel times all decrease. This demonstrates that the left-turning behavior
usually has the biggest impact on traffic coordination since left-turning CAVs
cross the conflict zone diagonally and are more likely to conflict with other
CAVs. In addition, it is worth noting that going straight produces the largest
travel time since this involves the largest number of MPs. However, when we
use the OCBF controller, the travel times under all situations are similar,
which shows that this controller can utilize the space resources of the
conflict zone and handle the influence of turns more effectively .

\subsection{Comparison results for more complicated intersections}

In this experiment, we consider more complicated intersections with two lanes
in each direction as shown in Fig. \ref{fig:inter}. The left lane in each
direction only allows going straight and turning left, while the right lane
only allows going straight and turning right. We set the arrival rate at all
lanes to be the same, i.e., 180 veh/h/lane and disallow lane-changing. Each
new incoming CAV chooses its behavior from the allowable behaviors with the
same probability, e.g., the CAV arriving at the entry of the left lane can go
straight or turn left with probability $0.5$. The comparison results are shown
in Table \ref{tab:table3}.

\begin{table}[ptbh]
\caption{Comparison results for a two-lane intersection}
\centering
\begin{tabular}
[c]{cccccc}%
\toprule $\beta$ & Methods & Energy & Travel time (s) & Fuel (mL) & Ave.
obj.\\
\midrule \multirow{3}[2]{*}{0.1} & SUMO & 24.0124 & 29.5955 & 30.5588 &
26.9720\\
& OC & 0.1514 & 28.5711 & 14.6685 & 3.0085\\
& OCBF & 1.0350 & 25.0804 & 18.5086 & 3.5430\\
\midrule \multirow{3}[2]{*}{0.5} & SUMO & 24.0124 & 29.5955 & 30.5588 &
38.8102\\
& OC & 0.6722 & 26.1284 & 17.2866 & 13.7364\\
& OCBF & 2.2244 & 22.6351 & 19.0753 & 13.5420\\
\midrule \multirow{3}[2]{*}{1} & SUMO & 24.0124 & 29.5955 & 30.5588 &
53.6079\\
& OC & 0.9063 & 25.8000 & 17.2933 & 26.7063\\
& OCBF & 2.9955 & 22.2347 & 19.1126 & 25.2302\\
\midrule \multirow{3}[2]{*}{2} & SUMO & 24.0124 & 29.5955 & 30.5588 &
83.2034\\
& OC & 1.2294 & 25.5773 & 17.3042 & 52.3840\\
& OCBF & 4.2353 & 22.1167 & 19.1500 & 48.4687\\
\bottomrule &  &  &  &  &
\end{tabular}
\label{tab:table3}%
\end{table}

The results here are similar to those in the single-lane intersections.
Although the number of MPs increases with the number of lanes, our method can
still effectively ensure safety and reduce travel time. It is worth noting
that the values of safety time headway for the OC controller are difficult to
determine. The OC controller requires that no CAV can enter the conflict
zone until the conflict CAV leaves it. However, the time spent for passing
through the conflict zone differs from vehicle to vehicle. If we choose a
larger value, then this significantly increases travel time and amplifies
conservatism. In contrast, if we choose a smaller value, the potential of
collision increases. Therefore, the MP-based method is significantly better
since it not only ensures safety but also reduces conservatism.

Next, we consider the impact of lane-changing on our OCBF method. For the same
two-lane intersection, we allow lane-changing and CAVs can choose any actions
(going straight, turning left and right). Since the left lane only allows
going straight and turning left, the right-turning CAV in this lane must
change its lane. The situation is similar for the left-turning CAV in the
right lane. To make a better comparison with the scenario without
lane-changing, we use the same arrival data (including the times all CAVs
enter the CZ and initial velocities) as the last experiment and only change
the lane that the turning CAV arrives at. For example, the left-turning CAVs
must arrive at the left lane in the last experiment, but in this experiment,
the lane they enter into can be random. The results are shown in Table
\ref{tab:lanechange}.

\begin{table}[ptbh]
\caption{The influence of lane-changing behaviour on the proposed method.}%
\label{tab:lanechange}
\centering
\begin{tabular}
[c]{cccccc}%
\toprule $\beta$ & Methods & Energy & Travel time & Fuel & Ave. obj.\\
\midrule \multirow{3}[2]{*}{0.1} & SUMO with LC & 23.9988 & 30.0337 &
30.0000 & 27.0022\\
& OCBF w/o LC & 1.0350 & 25.0804 & 18.5086 & 3.5430\\
& OCBF with LC & 1.0738 & 25.1200 & 18.5474 & 3.5858\\
\midrule \multirow{3}[2]{*}{0.5} & SUMO with LC & 23.9988 & 30.0337 &
30.0000 & 39.0157\\
& OCBF w/o LC & 2.2244 & 22.6351 & 19.0753 & 13.5420\\
& OCBF with LC & 2.2584 & 22.6689 & 19.1148 & 13.5929\\
\midrule \multirow{3}[2]{*}{1} & SUMO with LC & 23.9988 & 30.0337 & 30.0000 &
54.0325\\
& OCBF w/o LC & 2.9955 & 22.2347 & 19.1126 & 25.2302\\
& OCBF with LC & 3.0282 & 22.2684 & 19.1575 & 25.2966\\
\midrule \multirow{3}[2]{*}{2} & SUMO with LC & 23.9988 & 30.0337 & 30.0000 &
84.0662\\
& OCBF w/o LC & 4.2353 & 22.1167 & 19.1500 & 48.4687\\
& OCBF with LC & 4.2887 & 22.1536 & 19.2457 & 48.5959\\
\bottomrule &  &  &  &  &
\end{tabular}
\end{table}


We can see that the lane-changing behavior slightly increases all performance
measures compared with the results in scenarios disallowing lane-changing.
This is expected since a new (floating) MP is added and more control is
required to ensure safety. Nevertheless, the changes are small, fully
demonstrating the effectiveness of our method in handling lane-changing.
Although we have assumed that lane-changing only induces a fixed length, we
can extend our OCBF method with more complicated lane-changing
trajectories, e.g., trajectories fitted by polynomial functions. Note that in the SUMO simulation, it is assumed that a vehicle
can jump directly from one lane to another. However, our method still
outperforms it in all metrics, further supporting the advantages of the OCBF method.

Moreover, we explore the effect of asymmetrical arrival rates through two scenarios, in order to confirm that our OCBF method is effective even when traffic flows are heavy. In the first scenario, we set the arrival rates in Lanes 1, 2 to be three times as large as Lanes 3-8; while in the second scenario, the arrival rates in Lanes 1, 2, 5, 6 are three times as large as the remaining lanes. The comparison results are shown in Table \ref{tab:asymmetricalArrivalRates}.
\begin{table}[htbp]
  \centering
  \begin{threeparttable}
  \caption{The influence of asymmetrical and heavy traffic flows}
    \begin{tabular}{cccccc}
    \toprule
    Sce. & Methods  & Energy & Travel time & Fuel  & Ave. obj. \\
    \midrule
    \multirow{2}[2]{*}{1} & SUMO     & 43.1656 & 68.6488 & 40.4731 & 111.8144 \\
          & OCBF     & 2.6312 & 22.1812 & 19.1493 & 24.8124 \\
    \midrule
    \multirow{2}[2]{*}{2} & SUMO    & 44.4815 & 96.0517 & 44.6265 & 140.5332 \\
          & OCBF     & 3.0999 & 22.5594 & 19.5380 & 25.6593 \\
    \bottomrule
    \end{tabular}%
  \label{tab:asymmetricalArrivalRates}%
  \begin{tablenotes}
\footnotesize
\item[1] Scenario 1: the arrival rates at $l_1$ and $l_2$ are 540 veh/h/lane; while at $l_3$ to $l_8$ are 180 veh/h/lane.
\item[2] Scenario 2: the arrival rates at $l_1$, $l_2$, $l_5$, and $l_6$ are 540 veh/h/lane; while at $l_3$, $l_4$, $l_7$, and $l_8$ are 180 veh/h/lane.
\end{tablenotes}
\end{threeparttable}
\end{table}%

We can see in the SUMO simulation that traffic flows in lanes with high arrival rates are highly congested with CAVs forming long queues in these lanes. All metrics obtained from SUMO significantly increase compared with the results obtained from medium traffic shown in Table \ref{tab:table3}. However, since the coordination performance of our OCBF is much better than SUMO, all metrics remain at low levels, indicating the effectiveness of our OCBF controller in congested situations.

\subsection{The inclusion of the DR method}

Thus far in our experiments, the FIFO-based queue determines passing priority
when potential conflicts happen. This experiment extends our method to combine
it with a typical resequencing method, the DR method
\cite{zhang2018decentralized}. When a new CAV enters the CZ, the DR method
inserts it into the optimal position of the original crossing sequence. Note
that if we combine resequencing methods with the OC method, we need to update
the arrival times and trajectories of some CAVs whenever we adjust the
original crossing sequence. However, in the OCBF method, CAV $i$ only needs to
update its conflicting CAVs according to the new DR-based queue and follow the
original unconstrained optimal trajectory without replanning. In the following
experiments, we set $\beta=5$ and vary the length of some lanes to generate
different scenarios. The comparison results are shown in Table \ref{tab:DR}.

\begin{table}[ptbh]
\centering
\begin{threeparttable}
\caption{The effect of the DR method on the OCBF method}
\begin{tabular}{cccccc}
\toprule
Sce. & Methods & Energy & Travel time & Fuel  & Ave. obj. \\
\midrule
\multirow{2}[2]{*}{1} & OCBF+FIFO & 5.1261 & 21.6973 & 19.1518 & 113.6126 \\
& OCBF+DR & 5.1439 & 21.6404 & 18.9364 & 113.3459 \\
\midrule
\multirow{2}[2]{*}{2} & OCBF+FIFO & 5.9218 & 20.8093 & 18.6319 & 109.9683 \\
& OCBF+DR & 6.1080 & 20.6102 & 18.9077 & 109.1590 \\
\midrule
\multirow{2}[2]{*}{3} & OCBF+FIFO & 8.9344 & 19.3548 & 17.7396 & 105.7084 \\
& OCBF+DR & 7.0501 & 17.3253 & 17.2104 & 93.6766 \\
\bottomrule
\end{tabular}%
\label{tab:DR}
\begin{tablenotes}
\footnotesize
\item[1] Scenario 1 is a symmetrical intersection with all lanes are 300$m$.
\item[2] Scenario 2 is an asymmetrical Intersection with lane 3 and 4 are 200$m$ while lane 1, 2, 5, 6, 7, and 8 are 300$m$.
\item[3] Scenario 3 is an asymmetrical Intersection with lane 3 and 4 are 200$m$, lane 5 and 6 are 100$m$, while lane 1, 2, 7, and 8 are 300$m$.
\end{tablenotes}
\end{threeparttable}
\end{table}

The DR method helps decrease travel time and achieves a better average
objective value at the expense of energy consumption since CAVs need to do
more acceleration/deceleration maneuvers to adjust their crossing sequences.
The benefits of the DR method are more significant than the FIFO method for
asymmetrical intersections. This is because the FIFO rule may require a CAV
that enters the CZ later but is much closer to the intersection to yield to a
CAV that is far away from the intersection. For example, in the above Scenario
3, a CAV enters the CZ from lane 5 that is 100 m away from the intersection.
It is unreasonable to force it to yield to a CAV entering earlier but at 250 m
away from the intersection. Our method with resequencing can effectively avoid
producing such situations by replanning crossing sequences in an event-driven
way. Note that the OCBF+DR method outperforms the OCBF+FIFO method in all
metrics in Scenario 3 since nearly all CAVs arriving at lanes 5 and 6 need to
decelerate and even stop due to the FIFO rule, indicating that the DR method
is better than the simple FIFO principle when an intersection geometric
configuration is asymmetrical.

\subsection{The influence of nonlinear vehicle dynamics and noise}

We first consider the nonlinear vehicle dynamics in
(\ref{VehicleDynamicsNonlinear}) and reformulate all CBF constraints according
to the new dynamics. For the symmetrical intersection with two lanes in each
direction, we vary $\beta$ from 0.1 to 2 and use the OCBF+FIFO method to
coordinate the movements of CAVs. The results are shown in Table
\ref{tab:nonlinearVehicleDynamicsResults}.

\begin{table}[ptbh]
\caption{The influence of the nonlinear vehicle dynamics on the OCBF method}%
\label{tab:nonlinearVehicleDynamicsResults}
\centering
\begin{tabular}
[c]{ccccc}%
\toprule $\beta$ & Energy & Travel time & Fuel & Ave. obj.\\
\midrule 0.1 & 0.4751 & 24.5680 & 16.6496 & 2.9319\\
0.5 & 1.6822 & 22.3889 & 18.2822 & 12.8767\\
1 & 2.4702 & 21.9973 & 18.4682 & 24.4675\\
2 & 3.4667 & 21.7871 & 18.5832 & 47.0409\\
\bottomrule &  &  &  &
\end{tabular}
\end{table}

It is clear that the results conform to the results for the double integrator
vehicle dynamics (\ref{VehicleDynamics}). When $\beta$ increases, we are more
concerned about the travel time, thus travel time decreases while the energy
and fuel consumption rise. Note that though the nonlinear vehicle dynamics are
more complicated than the double integrator vehicle dynamics, the only
necessary modification is to derive the CBF constraints based on the new
dynamics. The computation times for these two different dynamics are nearly
the same.

Next, we have considered both noise and nonlinear dynamics. Due to the
measurement errors of sensors and imperfect actuators, there exists random
noise in position, velocity, and control inputs. To analyze the influence of
noise to the OCBF method, we consider uniformly distributed noise processes
($w_{i,p}(t)$ for the position of CAV $i$, $w_{i,v}(t)$ for the velocity, and
$w_{i,u}(t)$ for the control inputs) for this simulation. We set $\beta=0.1$
and use the OCBF+FIFO method for all experiments. The results are shown in
Table \ref{tab:noise}.

\begin{table}[ptbh]
\caption{The influence of noise on the OCBF method}%
\label{tab:noise}
\centering
\begin{tabular}
[c]{ccccc}%
\toprule Noise & Energy & Travel time & Fuel & Ave. obj.\\
\midrule no noise & 0.4751 & 24.5680 & 16.6496 & 2.9319\\
\midrule $w_{i,u}(t) \in[-0.5, 0.5]$ & 0.5777 & 24.6067 & 16.8714 & 3.0384\\
\midrule $w_{i,p}(t) \in[-1, 1]$ & \multirow{2}[2]{*}{5.4662} &
\multirow{2}[2]{*}{24.8587} & \multirow{2}[2]{*}{22.3935} &
\multirow{2}[2]{*}{7.9521}\\
$w_{i,v}(t) \in[-1, 1]$ &  &  &  & \\
\midrule $w_{i,p}(t) \in[-1, 1]$ & \multirow{3}[2]{*}{5.5723} &
\multirow{3}[2]{*}{24.8458} & \multirow{3}[2]{*}{22.3933} &
\multirow{3}[2]{*}{8.0569}\\
$w_{i,v}(t) \in[-1, 1]$ &  &  &  & \\
$w_{i,u}(t) \in[-0.5, 0.5]$ &  &  &  & \\
\midrule $w_{i,p}(t) \in[-2, 2]$ & \multirow{3}[6]{*}{31.3250} &
\multirow{3}[6]{*}{24.5667} & \multirow{3}[6]{*}{34.1352} &
\multirow{3}[6]{*}{33.7817}\\
$w_{i,v}(t) \in[-2, 2]$ &  &  &  & \\
$w_{i,u}(t) \in[-0.5, 0.5]$ &  &  &  & \\
\bottomrule &  &  &  &
\end{tabular}
\end{table}

The results show that the measurement errors of positions and velocities
significantly increase energy consumption. This is because noise causes CAVs
to misjudge their states necessitating additional control actions. For example,
suppose CAV $i$ is following CAV $j$ and their actual distance is 10 $m$ at
some time point, but, due to noise, CAV $i$ may misjudge this distance to be
8 $m$, therefore decelerating to enlarge their relative spacing. Then, at the
next time point, it may accelerate to keep a desired inter-vehicle space.
These frequent acceleration/deceleration maneuvers cause a considerable waste
of energy. As uncertainty increases, more control effort is needed to ensure
the safety of CAVs when the number of noise sources increases and the noise
magnitudes goes up. Note that CAVs may even collide with other CAVs when we
continuously increase the magnitude of noise. However, when noise is limited,
our method can effectively handle it and does not add any computational burden.

\section{CONCLUSIONS}

This paper presents a decentralized optimal control method for controlling
CAVs passing through a multi-lane intersection safely while jointly minimizing
the travel time and energy consumption of each CAV. First, CAVs calculate the
desired trajectory generated by unconstrained optimal control. Then, we design a search algorithm for a CAV to search all lateral safety constraints that it needs to meet with other conflicting CAVs. An OCBF controller is designed to follow the
desired trajectory while ensuring all safety constraints and physical vehicle
limitations. 
Multiple simulation experiments we have conducted show that the proposed method can handle complex objective functions, nonlinear vehicle dynamics, the presence of noise, and it is still effective under the influence of lane-changing behavior, heavy traffic flows, and asymmetrical intersections. Our ongoing work includes extensions to large traffic networks and enhancing the feasibility of the QPs in challenging environments where this may not always be the case. Finally, when the CAV trajectories involve curves, we plan to include more complicate lateral dynamics that better model vehicle behavior in such cases.

\addtolength{\textheight}{-12cm}




\bibliographystyle{IEEEtran}
\bibliography{IEEEabrv,reference}

\begin{thebibliography}{10}
\providecommand{\url}[1]{#1}
\csname url@samestyle\endcsname
\providecommand{\newblock}{\relax}
\providecommand{\bibinfo}[2]{#2}
\providecommand{\BIBentrySTDinterwordspacing}{\spaceskip=0pt\relax}
\providecommand{\BIBentryALTinterwordstretchfactor}{4}
\providecommand{\BIBentryALTinterwordspacing}{\spaceskip=\fontdimen2\font plus
\BIBentryALTinterwordstretchfactor\fontdimen3\font minus
  \fontdimen4\font\relax}
\providecommand{\BIBforeignlanguage}[2]{{%
\expandafter\ifx\csname l@#1\endcsname\relax
\typeout{** WARNING: IEEEtran.bst: No hyphenation pattern has been}%
\typeout{** loaded for the language `#1'. Using the pattern for}%
\typeout{** the default language instead.}%
\else
\language=\csname l@#1\endcsname
\fi
#2}}
\providecommand{\BIBdecl}{\relax}
\BIBdecl

\bibitem{rios2016survey}
J.~Rios-Torres and A.~A. Malikopoulos, ``A survey on the coordination of
  connected and automated vehicles at intersections and merging at highway
  on-ramps,'' \emph{IEEE Transactions on Intelligent Transportation Systems},
  vol.~18, no.~5, pp. 1066--1077, 2016.

\bibitem{chen2015cooperative}
L.~Chen and C.~Englund, ``Cooperative intersection management: A survey,''
  \emph{IEEE Transactions on Intelligent Transportation Systems}, vol.~17,
  no.~2, pp. 570--586, 2015.

\bibitem{li2014survey}
L.~Li, D.~Wen, and D.~Yao, ``A survey of traffic control with vehicular
  communications,'' \emph{IEEE Transactions on Intelligent Transportation
  Systems}, vol.~15, no.~1, pp. 425--432, 2014.

\bibitem{hult2016coordination}
R.~Hult, G.~R. Campos, E.~Steinmetz, L.~Hammarstrand, P.~Falcone, and
  H.~Wymeersch, ``Coordination of cooperative autonomous vehicles: Toward safer
  and more efficient road transportation,'' \emph{IEEE Signal Processing
  Magazine}, vol.~33, no.~6, pp. 74--84, 2016.

\bibitem{fayazi2018mixed}
S.~A. Fayazi and A.~Vahidi, ``Mixed-integer linear programming for optimal
  scheduling of autonomous vehicle intersection crossing,'' \emph{IEEE
  Transactions on Intelligent Vehicles}, vol.~3, no.~3, pp. 287--299, 2018.

\bibitem{xu2019grouping}
H.~Xu, S.~Feng, Y.~Zhang, and L.~Li, ``A grouping-based cooperative driving
  strategy for cavs merging problems,'' \emph{IEEE Transactions on Vehicular
  Technology}, vol.~68, no.~6, pp. 6125--6136, 2019.

\bibitem{qian2015decentralized}
X.~Qian, J.~Gregoire, A.~De~La~Fortelle, and F.~Moutarde, ``Decentralized model
  predictive control for smooth coordination of automated vehicles at
  intersection,'' in \emph{2015 European Control Conference (ECC)}.\hskip 1em
  plus 0.5em minus 0.4em\relax IEEE, 2015, pp. 3452--3458.

\bibitem{xu2020bi}
H.~Xu, Y.~Zhang, C.~G. Cassandras, L.~Li, and S.~Feng, ``A bi-level cooperative
  driving strategy allowing lane changes,'' \emph{Transportation Research Part
  C: Emerging Technologies}, vol. 120, p. 102773, 2020.

\bibitem{dresner2008multiagent}
K.~Dresner and P.~Stone, ``A multiagent approach to autonomous intersection
  management,'' \emph{Journal of artificial intelligence research}, vol.~31,
  pp. 591--656, 2008.

\bibitem{malikopoulos2018decentralized}
A.~A. Malikopoulos, C.~G. Cassandras, and Y.~J. Zhang, ``A decentralized
  energy-optimal control framework for connected automated vehicles at
  signal-free intersections,'' \emph{Automatica}, vol.~93, pp. 244--256, 2018.

\bibitem{zhang2019decentralized}
Y.~Zhang and C.~G. Cassandras, ``Decentralized optimal control of connected
  automated vehicles at signal-free intersections including comfort-constrained
  turns and safety guarantees,'' \emph{Automatica}, vol. 109, pp. 1--9, 2019.

\bibitem{meng2018analysis}
Y.~Meng, L.~Li, F.~Wang, K.~Li, and Z.~Li, ``Analysis of cooperative driving
  strategies for nonsignalized intersections,'' \emph{IEEE Transactions on
  Vehicular Technology}, vol.~67, no.~4, pp. 2900--2911, 2018.

\bibitem{zhang2018decentralized}
Y.~Zhang and C.~G. Cassandras, ``A decentralized optimal control framework for
  connected automated vehicles at urban intersections with dynamic
  resequencing,'' in \emph{2018 IEEE Conference on Decision and Control
  (CDC)}.\hskip 1em plus 0.5em minus 0.4em\relax IEEE, 2018, pp. 217--222.

\bibitem{xu2019cooperative}
\BIBentryALTinterwordspacing
H.~Xu, Y.~Zhang, L.~Li, and W.~Li, ``Cooperative driving at unsignalized
  intersections using tree search,'' \emph{IEEE Transactions on Intelligent
  Transportation Systems}, pp. 1--9, 2019. [Online]. Available:
  \url{https://ieeexplore.ieee.org/abstract/document/8848467}
\BIBentrySTDinterwordspacing

\bibitem{xiao2019decentralized}
W.~Xiao and C.~G. Cassandras, ``Decentralized optimal merging control for
  connected and automated vehicles,'' in \emph{2019 American Control Conference
  (ACC)}.\hskip 1em plus 0.5em minus 0.4em\relax IEEE, 2019, pp. 3315--3320.

\bibitem{Wei2020}
W.~Xiao, C.~G. Cassandras, and C.~Belta, ``Bridging the gap between optimal
  trajectory planning and safety-critical control with applications to
  autonomous vehicles,'' \emph{Automatica (provisionally accepted, available in
  arXiv: 2008.07632)}, 2020.

\bibitem{ames2014control}
A.~D. Ames, J.~W. Grizzle, and P.~Tabuada, ``Control barrier function based
  quadratic programs with application to adaptive cruise control,'' in
  \emph{53rd IEEE Conference on Decision and Control (CDC)}.\hskip 1em plus
  0.5em minus 0.4em\relax IEEE, 2014, pp. 6271--6278.

\bibitem{ames2019control}
A.~D. Ames, S.~Coogan, M.~Egerstedt, G.~Notomista, K.~Sreenath, and P.~Tabuada,
  ``Control barrier functions: Theory and applications,'' in \emph{2019 18th
  European Control Conference (ECC)}.\hskip 1em plus 0.5em minus 0.4em\relax
  IEEE, 2019, pp. 3420--3431.

\bibitem{Xiao2019}
W.~Xiao and C.~Belta, ``Control barrier functions for systems with high
  relative degree,'' in \emph{Proc. of 58th IEEE Conference on Decision and
  Control}, Nice, France, 2019, pp. 474--479.

\bibitem{Wei2020ACC}
W.~Xiao and C.~G. Cassandras, ``Decentralized optimal merging control for
  connected and automated vehicles with optimal dynamic resequencing,'' in
  \emph{Proc. of the American Control Conference}, 2020, pp. 4090--4095.

\bibitem{vogel2003comparison}
K.~Vogel, ``A comparison of headway and time to collision as safety
  indicators,'' \emph{Accident analysis \& prevention}, vol.~35, no.~3, pp.
  427--433, 2003.

\bibitem{khalil2002nonlinear}
H.~K. Khalil and J.~W. Grizzle, \emph{Nonlinear systems}.\hskip 1em plus 0.5em
  minus 0.4em\relax Prentice hall Upper Saddle River, NJ, 2002, vol.~3.

\bibitem{xiao2019decentralizedITSC}
W.~Xiao, C.~G. Cassandras, and C.~Belta, ``Decentralized merging control in
  traffic networks with noisy vehicle dynamics: a joint optimal control and
  barrier function approach,'' in \emph{2019 IEEE Intelligent Transportation
  Systems Conference (ITSC)}.\hskip 1em plus 0.5em minus 0.4em\relax IEEE,
  2019, pp. 3162--3167.

\bibitem{Wei2019}
W.~Xiao, C.~Belta, and C.~G. Cassandras, ``Decentralized merging control in
  traffic networks: A control barrier function approach,'' in \emph{Proc.
  ACM/IEEE International Conference on Cyber-Physical Systems}, Montreal,
  Canada, 2019, pp. 270--279.

\bibitem{kamal2012model}
M.~A.~S. Kamal, M.~Mukai, J.~Murata, and T.~Kawabe, ``Model predictive control
  of vehicles on urban roads for improved fuel economy,'' \emph{IEEE
  Transactions on control systems technology}, vol.~21, no.~3, pp. 831--841,
  2012.

\end{thebibliography}

\end{document}